\documentclass[floatfix,reprint,amsmath,amssymb,aps,pra]{revtex4-1}
\usepackage{times,mathptmx}
\usepackage{graphicx}
\usepackage[english]{babel}
\usepackage{epstopdf}
\usepackage{longtable}
\usepackage{xcolor}

\begin{document}

\title{Quasi-bound states in the continuum induced by $\mathcal{PT}$-symmetry breaking}

\author{Denis~V.~Novitsky$^{1}$}
\email{dvnovitsky@gmail.com}
\author{Alexander~S.~Shalin$^{2,3,4}$}
\author{Dmitrii Redka$^{5,4}$}
\author{Vjaceslavs Bobrovs$^4$}
\author{Andrey~V.~Novitsky$^{6,2}$}

\affiliation{$^1$B.~I.~Stepanov Institute of
Physics, National Academy of Sciences of Belarus, Nezavisimosti
Avenue 68, 220072 Minsk, Belarus \\ 
$^2$ITMO University, Kronverksky Prospekt 49, 197101 St. Petersburg, Russia \\
$^3$Kotel’nikov Institute of Radio Engineering and Electronics, Russian Academy of Sciences (Ulyanovsk branch), Goncharova Str. 48, 432000 Ulyanovsk, Russia \\
$^4$Riga Technical University, Institute of Telecommunications, Azenes st. 12, 1048 Riga, Latvia \\
$^5$Saint Petersburg Electrotechnical University “LETI” (ETU),  Prof. Popova Street 5, 197376 St. Petersburg, Russia \\
$^6$Department of Theoretical Physics and Astrophysics, Belarusian State University, Nezavisimosti Avenue 4, 220030 Minsk, Belarus}

\date{\today}

\begin{abstract}
Bound states in the continuum (BICs) enable unique features in tailoring light-matter interaction on nanoscale. These radiationless localized states drive theoretically infinite quality factors and lifetimes for modern nanophotonics, making room for a variety of emerging applications. Here we use the peculiar properties possessed by the so-called $\mathcal{PT}$-symmetric optical structures to propose the novel mechanism for the quasi-BIC manifestation governed by the $\mathcal{PT}$-symmetry breaking. In particular, we study regularities of the spontaneous $\mathcal{PT}$-symmetry breaking in trilayer structures with the outer loss and gain layers consisting of materials with permittivity close to zero. We reveal singular points on the curves separating $\mathcal{PT}$-symmetric and broken-$\mathcal{PT}$-symmetry states in the parametric space of the light frequency and the angle of incidence. These singularities remarkably coincide with the BIC positions at the frequency of volume plasmon excitation, where the dielectric permittivity vanishes. The loss and gain value acts as an asymmetry parameter that disturbs conditions of the ideal BIC inducing the quasi-BIC. Fascinating properties of these quasi-BICs having ultrahigh quality factors and almost perfect transmission can be utilized in sensing, nonlinear optics, and other applications.
\end{abstract}

\maketitle

\section{Introduction}

Photonic systems with balanced loss and gain, also known as $\mathcal{PT}$-symmetric systems, have attracted much attention over the last decade. Being at first a mere classical analogue of a peculiar quantum-mechanical invariance \cite{Bender1998}, the $\mathcal{PT}$ symmetry was soon realized as a powerful tool to manipulate light-matter interaction \cite{Zyablovsky2014, Feng2017, El-Ganainy2018, Ozdemir2019}. A number of optical schemes were used to observe the $\mathcal{PT}$ symmetry including coupled waveguides \cite{Ruter2010}, photonic lattices \cite{Regensburger2012}, and two-dimensional crystals \cite{Kremer2019}. Perhaps, the simplest one is a layered structure with alternating loss and gain materials \cite{Ge2012}. Studies of $\mathcal{PT}$-symmetric systems initiated rapid development of non-Hermitian photonics devoted to general problems of open photonic systems changing researchers' attitude to loss and gain: loss is more than an attenuation factor to get rid of and gain is more than a means to reach amplification and lasing. As an example of this more general approach, asymmetric, unbalanced distributions of loss and gain have attracted attention recently, e.g., for loss compensation \cite{Hlushchenko2020} or lasing-threshold tuning \cite{Liang2021}.

One of the most intriguing features of the $\mathcal{PT}$ symmetry in various systems exhibits itself in its spontaneous breakdown. This can be achieved by tuning parameters of the system, for example, the level of balanced loss and gain. Under the tuning, one reaches a moment when a modal composition of the system degenerates and the $\mathcal{PT}$ symmetry gets broken. Such degeneracies called exceptional points (EPs) are important peculiarities of many optical and photonic systems and, generally, correspond to the simultaneous coalescence of complex eigenvalues and eigenvectors of non-Hermitian Hamiltonians \cite{Miri2019}. In particular, the EPs in $\mathcal{PT}$-symmetric structures are useful for observing enhanced sensing \cite{Chen2017, Hodaei2017, Yu2020}, single-mode lasing \cite{Feng2014-2, Hodaei2014}, coherent perfect absorption \cite{Longhi2010, Wong2016, Novitsky2019}, slow light \cite{Goldzak2018}, polarization-state conversion \cite{Hassan2017}, nonreciprocal transmission \cite{Novitsky2018}, topologically protected states \cite{Weimann2016, Song2020, Parto2021}, and so on. Note that the $\mathcal{PT}$-symmetry breaking at the EP can be treated in terms of transition between the $\mathcal{PT}$-symmetric phase and broken-$\mathcal{PT}$-symmetry phase being one more application of the concept of phase transitions in the laser and optical physics \cite{DeGiorgio1970, Xie2020}.

On the other hand, in the last decade much attention has been paid to the so-called bound states in the continuum (BICs) predicted first in quantum systems \cite{Wigner1929}. BICs can be considered as point features in reflection or scattering spectra of optical structures, arising, for example, when several resonant responses (modes) are superimposed \cite{Hsu2016, Azzam2021, Koshelev2020, Sadreev2021}. As a result of such superposition at a certain frequency and angle of light incidence, resonance states having ideally an infinitely high (divergent) quality ($Q$) factors can arise. Such states are characterized by perfect localization of radiation and, as a consequence, cannot be excited by light incident on the structure from outside, i.e., the BICs are dark (trapped) modes decoupled from the continuum of radiation. Although strict BICs are unobservable, slightly deviating from the ideal BIC conditions, it is possible to excite quasi-BICs arising in the form of very narrow (high-$Q$) Fano resonances. In the optical context, BICs (although not under this name) were first shown to exist in photonic crystal structures in Refs. \cite{Astratov1999, Paddon2000, Ochiai2001}. Subsequently, BICs have been observed or theoretically discussed in arrays of dielectric and metal-dielectric elements \cite{Marinica2008, Bulgakov2008, Bulgakov2017, Azzam2018}, waveguide structures \cite{Plotnik2011, Weimann2013, Gomis-Bresco2017}, photonic crystals \cite{Lee2012, Hsu2013}, metamaterials and metasurfaces \cite{Fedotov2007, Koshelev2018, Kupriianov2019}, and even single dielectric nanoparticles \cite{Bogdanov2019, Huang2021}. Potential applications of quasi-BICs include enhancement of nonlinear response \cite{Koshelev2020a, Bulgakov2019}, lasing \cite{Kodigala2017, Gongora2021}, generation of optical vortices \cite{Huang2020, BWang2020}, and sensing \cite{Maksimov2020, YWang2021}.

Peculiar BICs appear in non-Hermitian systems and their existence is tightly connected with the EPs. In particular, it was predicted that the BICs can be observed in the broken-$\mathcal{PT}$-symmetry regime in one- and two-dimensional waveguide systems with loss and gain \cite{Regensburger2013, Molina2014, Kartashov2018}. Another type of BICs was shown to exist in both $\mathcal{PT}$-symmetric and broken-$\mathcal{PT}$-symmetry phases within the coupled-waveguides framework \cite{Longhi2014a}. Specific BIC-like unstable states can exist at the EPs of non-Hermitian defective lattices \cite{Longhi2014b}. Unconventional BICs were reported to appear in the anti-$\mathcal{PT}$-symmetric phase in the cavity-magnonics systems \cite{Yang2020} and open quantum systems with $\mathcal{PT}$-symmetric defects \cite{Garmon2015}. However, the presence of the EP is not a necessary condition for BICs assisted by loss and gain, since the non-Hermiticity can be used for controlling coupling between resonances lying behind the quasi-BIC phenomenon \cite{Gandhi2020a, Gandhi2020b}. Finally, a novel mechanism for generation of the BICs outside the scope of the EP physics was proposed in the BIC-supporting systems under $\mathcal{PT}$-symmetric perturabtion \cite{Song2020}.

In this paper, starting with the system supporting a BIC in the Hermitian limit, we study how properties of the BIC alter when loss and gain are introduced. In particular, we consider the $\mathcal{PT}$-symmetric layered system containing an epsilon-near-zero (ENZ) material. It was shown recently that the systems with singular (ENZ-like) properties possess BICs caused by coupling between plasmonic and Fabry-Perot resonances \cite{Monticone2018}. In general, the BIC-supporting one-dimensional (layered) structures require either materials with singular properties or anisotropic media for mixing light of different polarizations \cite{Gomis-Bresco2017, Pankin2020}. Subsequently, the unique optical, thermal, and topological properties of the ENZ-related BICs have been studied in detail \cite{Duggan2019, Sakotic2020, Sakotic2021}. We put these BICs into the context of the $\mathcal{PT}$ symmetry research by adding balanced loss and gain. This was shown to lead to appearance of the quasi-BIC \cite{Sakotic2019}, but the specific mechanism of BIC transformation into quasi-BIC has not been revealed yet. Here, we fill in this gap and show that the high-$Q$ quasi-BIC resonances in the ENZ-containing $\mathcal{PT}$-symmetric layered systems are induced by the coincidence of the BIC with a singular point of the $\mathcal{PT}$-symmetry breaking phase diagram. The non-Hermiticity magnitude (loss and gain value) takes on the role of the structure asymmetry parameter leading to the peculiar quasi-BIC with symmetric line shape, perfect transmission and strong light localization inside the structure. We note that the distinction between the true BIC and quasi-BIC is important for our discussion, although the quasi-BICs can be often treated just as the BICs in many realistic situations. Thus, the $\mathcal{PT}$-symmetry-breaking singularity offers a novel mechanism behind the excitation of quasi-BICs extending the remarkable diversity of BIC physics known nowadays.

\begin{figure}[t!]
\centering \includegraphics[scale=1.0, clip=]{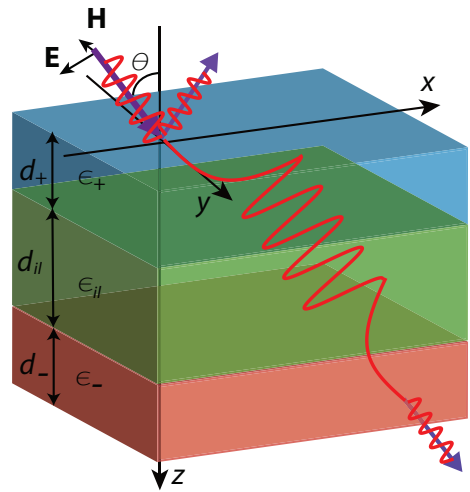}
\caption{\label{fig1} Schematic of a $\mathcal{PT}$-symmetric trilayer with outer layers containing lossy and gainy ENZ media and dielectric interlayer in between. The outer layers have the thickness $d_\pm=\lambda_p/2 \pi$ (i.e., $\omega_p d_\pm/c=1$, where $c$ is the speed of light); the interlayer has the thickness $d_{il}=10 d_\pm$ and permittivity $\varepsilon_{il}=5$.}
\end{figure}

\section{$\mathcal{PT}$-symmetric trilayers} 

We start with description of our $\mathcal{PT}$-symmetric system and the origin of phase transition there. The simplest $\mathcal{PT}$-symmetric layered structure is the bilayer one, which is a well-studied system, see, e.g., our recent analysis \cite{Novitsky2020}. The $\mathcal{PT}$-symmetric bilayer consists of just two layers -- one with loss (permittivity $\varepsilon_+$) and another with gain ($\varepsilon_-$). The $\mathcal{PT}$-symmetric trilayer has an additional interlayer (spacer) with the real-valued permittivity $\varepsilon_{il}$ located between the loss and gain side layers (see Fig. \ref{fig1}). Trilayers are much less studied and are in the spotlight of this paper. The introduction of the loss-free and gain-free interlayer dramatically changes the phase transition patterns of the system.

We study availability of phase transitions, that is $\mathcal{PT}$-symmetry breaking, in trilayers as a function of light-wave angle of incidence. The $\mathcal{PT}$-symmetry breaking phenomenon can be described in terms of the scattering matrix eigenvalues and eigenvectors. The scattering matrix of a multilayered structure has the form $\hat S = \left(
\begin{array}{cc} t & r_R \\ r_L & t \end{array} \right)
\label{scat}$, where $t$ is the transmission coefficient, $r_L$ and $r_R$ are the reflection coefficients for the left- and right-incident waves \cite{Novitsky2020}. Eigenvalues $s_{1,2}$ of the matrix $\hat S$ are known to be both unimodular ($|s_{1,2}|=1$) in the $\mathcal{PT}$-symmetric state and inversely proportional ($|s_{1}|=1/|s_{2}|$) in the broken-$\mathcal{PT}$-symmetry state. Point of the phase transition where behavior of the eigenvalues $s_{1,2}$ dramatically changes is the exceptional point. Eigenvectors of the scattering matrix $\hat S$ coincide at the EP, so that the system becomes degenerate there. Transmission and reflection coefficients used in the scattering matrix formulation can be calculated with the well-known transfer-matrix method. We limit our consideration to the TM polarization (see Appendix \ref{AppA} for details) to deal with plasmon excitation.

\begin{figure*}[t!]
\centering \includegraphics[scale=1, clip=]{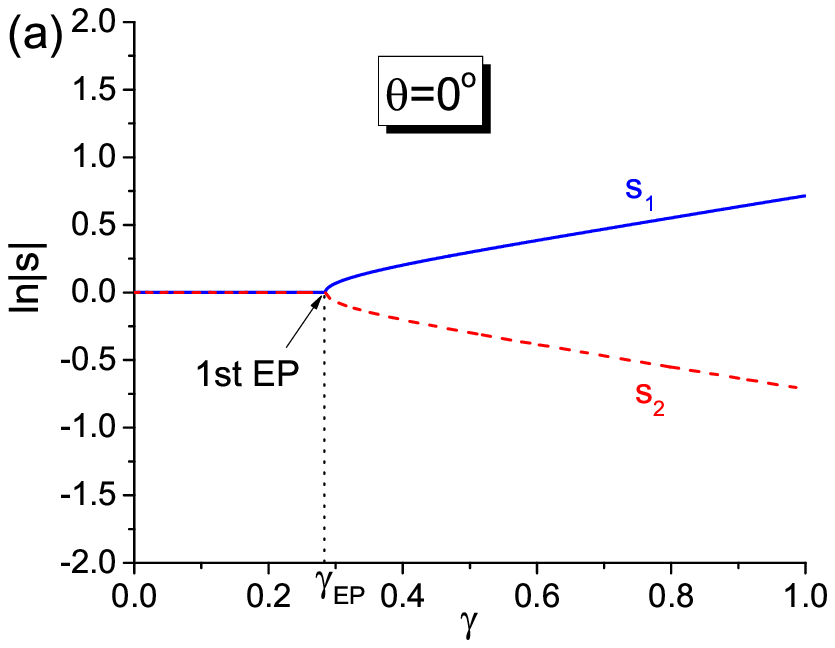}
\includegraphics[scale=1, clip=]{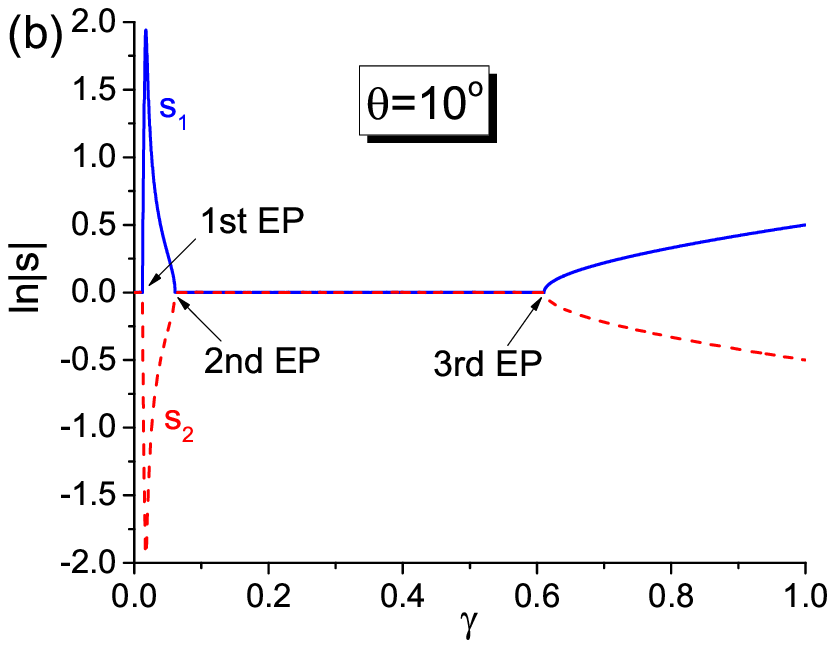}
\includegraphics[scale=1, clip=]{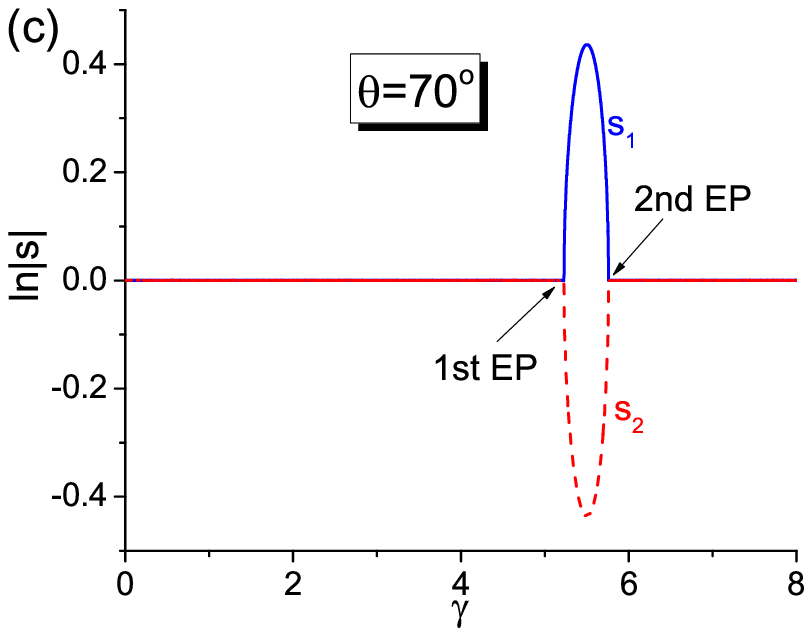}
\caption{\label{fig2} Dependence of the logarithm of the scattering-matrix eigenvalues on the loss and gain coefficient $\gamma$ for the $\mathcal{PT}$-symmetric trilayer. The light frequency is $\omega=\omega_p$; three different incidence angles are (a) $\theta=0^{\circ}$, (b) $\theta=10^{\circ}$, and (c) $\theta=70^{\circ}$. We employ parameters of the structure indicated in the caption of Fig. \ref{fig1}.}
\end{figure*}

We take the permittivities $\varepsilon_+$ and $\varepsilon_-$ of the loss and gain layers, respectively, as
\begin{equation}
\varepsilon_{\pm}=1 \pm i \gamma - \frac{\omega^2_p}{\omega^2},
\label{epslg}
\end{equation}
where $\gamma>0$ is the loss or gain coefficient (non-Hermiticity magnitude) and $\omega_p$ is the plasma frequency. We are interested in the ENZ regime observed in the vicinity of $\omega_p$, since in this case a trilayer could support a BIC \cite{Monticone2018}. There are several reasons, why we use Eq. (\ref{epslg}) instead of the standard Drude formula as in Ref. \cite{Monticone2018}. First, the permittivity (\ref{epslg}) allows one to separate the effects of ENZ and non-Hermiticity, so that the physical picture becomes as clear as possible. Second, Eq. (\ref{epslg}) can be obtained from the Drude formula $\varepsilon=1 - \omega^2_p/(\omega^2 + i \Gamma \omega)$ under $|\Gamma| \ll \omega$ and $\gamma \approx \Gamma \omega^2_p/\omega^3 \approx \Gamma/\omega_p$, when we are able to neglect dependence of $\gamma$ on the frequency considering a relatively narrow range near $\omega_p$. (See Appendix \ref{AppG} for calculations showing that the main results of this paper can be reproduced with the Drude dispersion as well.) Third, since we are interested in determining a dependence of the system response on the loss and gain level, the easiest way is just to vary $\gamma$ freely at a given frequency. Such variations are most comprehensible from Eq. (\ref{epslg}). Finally, if the first term of $\varepsilon_{\pm}$ took non-unit values, then it would result only in shifting the ENZ condition to a different frequency. The ENZ media needed can be realized either with the well-known Drude materials \cite{Kinsey2019}, such as metals and transparent conducting oxides (although introduction of gain in these materials is not always feasible in practice \cite{Sakotic2021}), or with a low-loss zero-index metamaterial for the ENZ component \cite{Monticone2018, Shalin2010, Terekhov2019a, Terekhov2019b} and the loss or gain material embedded in it. Thus, the choice of the permittivity in the form of Eq. (\ref{epslg}) does not limit the generality of our analysis and allows us to consider both ENZ and non-Hermiticity effects in a simple, convenient way. The thicknesses of the loss and gain layers are supposed to be the same, $d_+=d_-$; the interlayer is characterized with $\varepsilon_{il}$ and $d_{il}$; in the case of a bilayer, one should set $d_{il}=0$.

\section{Phase transitions in $\mathcal{PT}$-symmetric trilayers} 

\begin{figure}[t!]
\centering \includegraphics[scale=1, clip=]{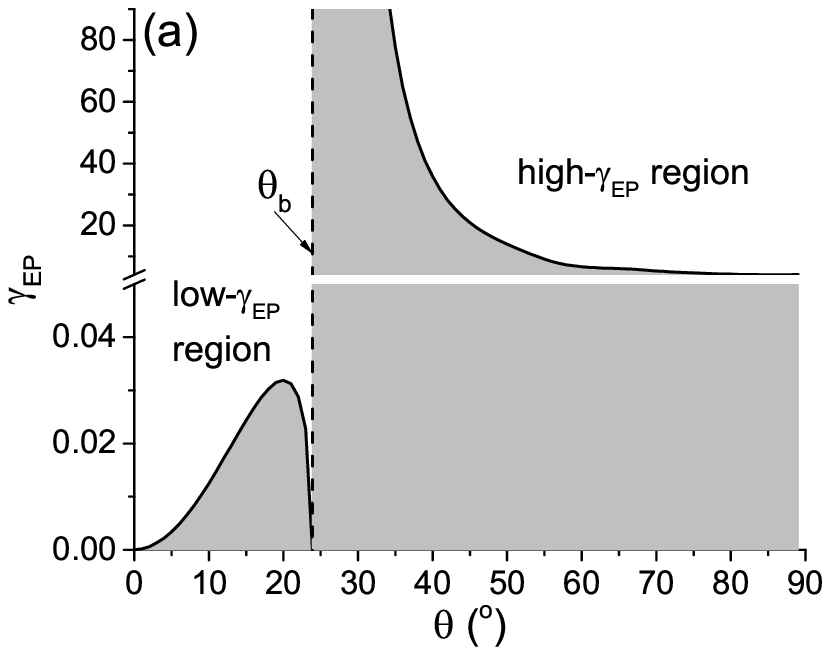}
\includegraphics[scale=1, clip=]{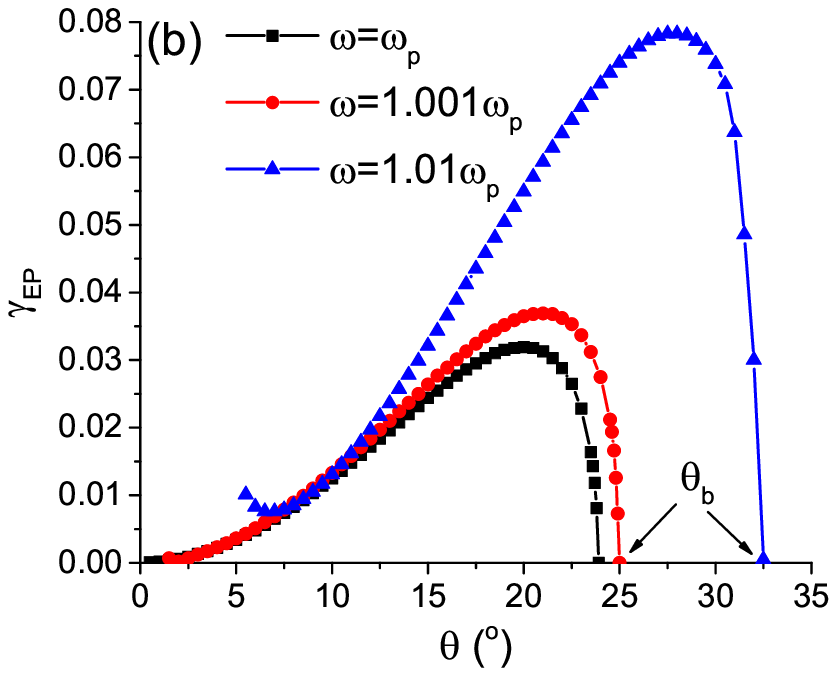}
\caption{\label{fig3} The EP loss and gain level, $\gamma_{EP}$, as a function of the angle of incidence for the $\mathcal{PT}$-symmetric trilayer. (a) Full angular dependence at $\omega=\omega_p$. The gray area corresponds to $\gamma<\gamma_{EP}$, where $\mathcal{PT}$ symmetry is preserved. (b) The low-$\gamma_{EP}$ region for the three different frequencies around $\omega_p$. Parameters of the structure are the same as in Fig. \ref{fig1}.}
\end{figure}

\begin{figure}[t!]
\centering \includegraphics[scale=1, clip=]{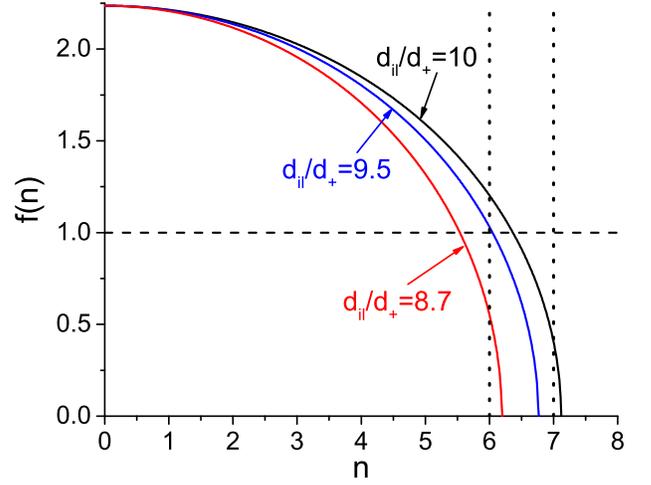}
\caption{\label{fig4} Plot of the function $f(n)=\sqrt{\varepsilon_{il}- \left( \frac{\pi c n}{\omega_p d_{il}} \right)^2}$ for different interlayer thicknesses. The other parameters of the structure are the same as in Fig. \ref{fig1}.}
\end{figure}

Let us study behavior of the minimal loss and gain levels corresponding to the first EP (we denote it as $\gamma_{EP}$) as a function of the angle of incidence $\theta$. Several examples of curves for scattering matrix eigenvalues $|s_{1}|$ and $|s_{2}|$ illustrating the $\mathcal{PT}$-symmetry breaking are shown in Fig. \ref{fig2}. The point $\gamma=\gamma_{EP}$, in which the curves for $|s_{1}|$ and $|s_{2}|$ diverge is called the exceptional point. One can see that at normal incidence, the $\mathcal{PT}$-symmetry breaking occurs for comparatively large non-Hermiticity magnitude, $\gamma_{EP} \approx 0.285$ [see Fig. \ref{fig2}(a)]. On the contrast, at $\theta=10^{\circ}$, the loss and gain needed for the EP are much lower, $\gamma_{EP} \approx 0.0125$ [see Fig. \ref{fig2}(b)]. Finally, at the large incident angle $\theta=70^{\circ}$, the non-Hermiticity magnitude needed for reaching the first EP gets much larger, $\gamma_{EP} \approx 5.2$ [see Fig. \ref{fig2}(c)]. One can also see in Fig. \ref{fig2} the second and third EPs at higher levels of the non-Hermiticity parameter. Further, we focus on the first EP and study the transition between the low-$\gamma_{EP}$ and high-$\gamma_{EP}$ regions. The behavior of other EPs is briefly discussed in Appendix \ref{AppB}.

In Fig. \ref{fig3}(a), the dependence $\gamma_{EP}(\theta)$ for the first EP is shown at $\omega=\omega_p$. We see that the full angular range is divided into two regions, where $\gamma_{EP}$ takes on either low or high values. The boundary between these regions denoted as $\theta_b$ is a peculiar singular point: As we reach $\theta_b$ from the left, $\gamma_{EP} \rightarrow 0$ (violation of the $\mathcal{PT}$ symmetry is easily reached), whereas $\gamma_{EP} \rightarrow \infty$ just above $\theta_b$ (the $\mathcal{PT}$ symmetry is never broken). Even finite, but large values of the loss and gain ($\gamma_{EP} > 4$) needed for breaking the $\mathcal{PT}$ symmetry at large angles $\theta > \theta_b$ make the phase transition hardly observable or even impossible. Thus, system's behavior strongly differs in the low-$\gamma_{EP}$ and high-$\gamma_{EP}$ regions.

Using the transfer matrix of the structure, the value of $\theta_b$ can be estimated analytically at $\omega=\omega_p$. Indeed, the singular point corresponds to the condition $|t| > 1$ for $\gamma \rightarrow \infty$, that is the $\mathcal{PT}$-symmetry breaking occurs only for very large loss and gain as we reach $\theta_b$ from the high-$\gamma_{EP}$ side. A simple estimate reads as follows
\begin{equation}
\theta_b (\omega=\omega_p) = \arcsin \sqrt{\varepsilon_{il}- \left( \frac{\pi c n}{\omega_p d_{il}} \right)^2},
\label{thetab}
\end{equation}
where $n$ is an integer number. Derivation of Eq. (\ref{thetab}) is discussed in Appendix \ref{AppC}. To clearly represent how this estimate can be used, in Fig. \ref{fig4} we plot the function $f(n)=\sqrt{\varepsilon_{il} - \left( \frac{\pi c n}{\omega_p d_{il}} \right)^2}$, where $n$ is assumed to be continuous. In the case of $d_{il}=10 d_\pm$ discussed in Fig. \ref{fig3}(a) there is a single discrete value $n=7$ satisfying the sine-value limitation, $0 \leq f(n) \leq 1$. So, for $n=7$, we obtain $\theta_b \approx 23.881^{\circ}$. which is in perfect agreement with numerical calculations shown in Fig. \ref{fig3}(a).

\begin{figure}[t!]
\centering \includegraphics[scale=1, clip=]{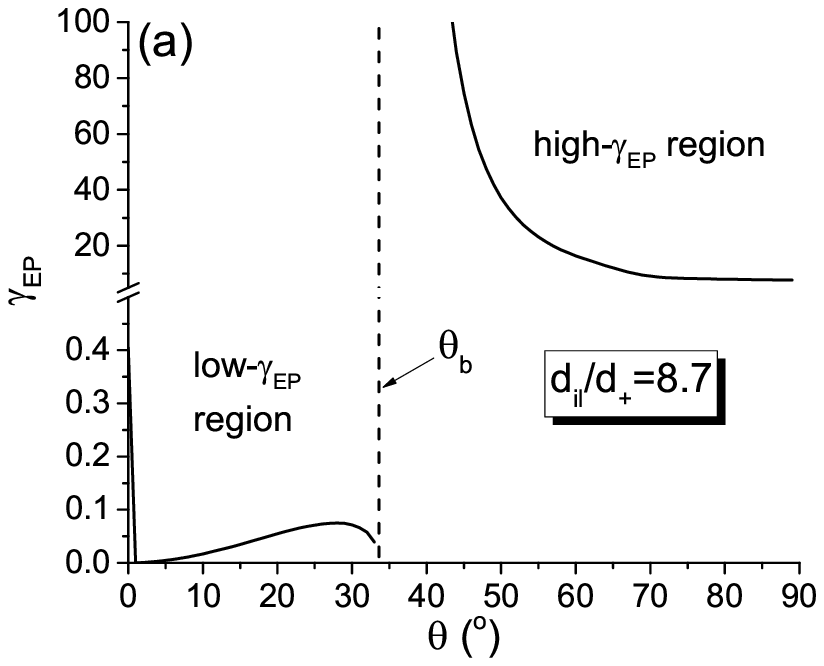}
\includegraphics[scale=1, clip=]{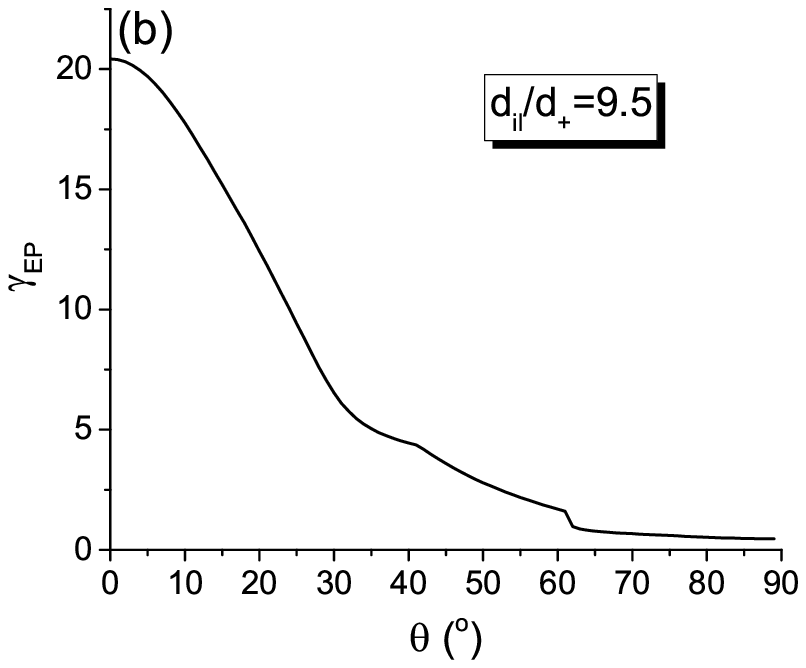}
\caption{\label{fig5} The EP loss and gain level, $\gamma_{EP}$, as a function of the incident angle for different interlayer thicknesses (a) $d_{il}=8.7 d_\pm$ and (b) $d_{il}=9.5 d_\pm$ of the $\mathcal{PT}$-symmetric trilayer. The other parameters of the structure are the same as in Fig. \ref{fig1}.}
\end{figure}

To illustrate that this approach works for other situations as well, we consider two other cases. For $d_{il}=8.7 d_\pm$, one should take $n=6$ to obtain $\theta_b \approx 33.571^{\circ}$. This is supported by numerical calculations of $\gamma_{EP}$ shown in Fig. \ref{fig5}(a): we again see the low-$\gamma_{EP}$ and high-$\gamma_{EP}$ regions below and above $\theta_b$. On the contrary, for $d_{il}=9.5 d_\pm$, there is no any suitable discrete $n$ satisfying $0 \leq f(n) \leq 1$. As a result, high- and low-$\gamma_{EP}$ regions are not available in this case [see Fig. \ref{fig5}(b)]: $\gamma_{EP}$ changes monotonously and there are no any breaks.

The dependencies $\gamma_{EP}(\theta)$ for several light frequencies in the low-$\gamma_{EP}$ region are demonstrated in Fig. \ref{fig3}(b). In order to make the figure more readable, we do not show the low-angle data with higher $\gamma_{EP}$ observed in Fig. \ref{fig3}(a). It is clear from Fig. \ref{fig3}(b) that the low-$\gamma_{EP}$ region gets wider and $\theta_b$ shifts to higher angles, when the frequency is above the plasma one. We would like also to draw attention to the sharp break of the curves near the singular point indicating a potential for significant modification of the structure response with a tender tuning of the angle of incidence.

\begin{figure}[t!]
\centering \includegraphics[scale=1, clip=]{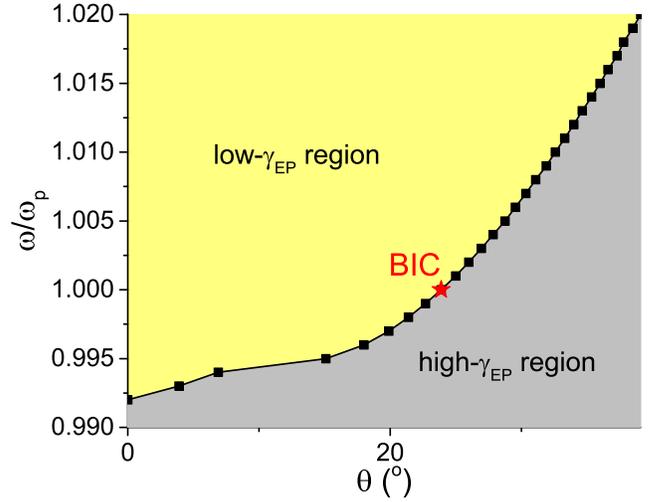}
\caption{\label{fig6} Phase diagram of the $\mathcal{PT}$-symmetric trilayer in the ``frequency -- angle'' coordinates. Parameters of the structure are the same as in Fig. \ref{fig1}.}
\end{figure}

As demonstrated in Fig. \ref{fig6}, the full ``frequency -- angle'' plane is divided into two regions possessing different behaviors. In the low(high)-$\gamma_{EP}$ region, the $\mathcal{PT}$-symmetry breaking is easily (hardly) achievable. A sharp boundary between the low- and high-$\gamma_{EP}$ regions consists of singular points, in which the value of $\gamma_{EP}$ is not determined. It is important that changing the permittivity and thickness of the interlayer, we can vary widely these regions in the phase diagram. The brief discussion of the interlayer thickness influence can be found in Appendix \ref{AppD}. Thus, the interlayer being just a lossless dielectric strongly influences the response of the $\mathcal{PT}$-symmetric trilayer structures, what can be useful in sensing applications.

\section{Quasi-BICs via the $\mathcal{PT}$-symmetry breaking} 

\begin{figure}[t!]
\centering \includegraphics[scale=1, clip=]{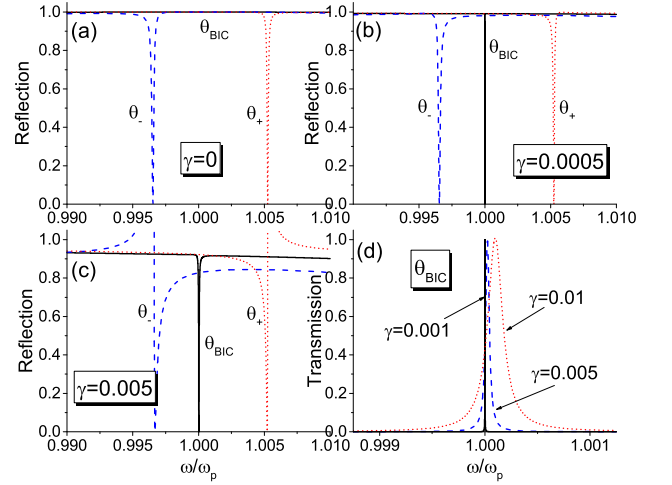}
\caption{\label{fig7} (a)-(c) Reflection and (d) transmission spectra of the $\mathcal{PT}$-symmetric trilayer for different loss and gain levels $\gamma$. The spectra for the incident angles $\theta_{BIC}$ and $\theta_\pm=\theta_{BIC} \pm 5^{\circ}$ are shown. The other parameters are the same as in Fig. \ref{fig1}.}
\end{figure}

Let us demonstrate how the singular point discussed above can be used for controlling quasi-BICs in ENZ-containing layered structures. It is known that a lossless trilayer possesses a BIC at the plasma frequency and a certain incident angle given by $\theta_{BIC} = \arcsin \sqrt{\varepsilon_{il}-(\pi c n / \omega_p d_{il})^2}$ with $n=0,1,2...$ \cite{Monticone2018}. This type of the BIC is the result of exact destructive interference between the narrowband volume-plasmon resonance in the ENZ layers and the broad Fabry-Perot resonance of the dielectric interlayer. When we detune from the BIC position, the imperfect interference manifests in the spectra as asymmetric Fano profiles. An example is shown in Fig. \ref{fig7}(a): the narrow dips in reflection of lossless structure ($\gamma=0$) appear as we departure from the BIC angle to $\theta_\pm=\theta_{BIC} \pm 5^{\circ}$.

One can see that the BIC given by the above expression coincides with the singular point Eq. (\ref{thetab}) of the phase diagram at the plasma frequency, i.e., $\theta_{BIC}=\theta_b$. This fact can be understood in terms of the scattering-matrix poles and zeros whose convergence gives rise to both BICs \cite{Monticone2018} and singular points of $\mathcal{PT}$-symmetric systems \cite{Chong2011, Krasnok2019}. It should be stressed that the convergence for BICs and EP singularities has different nature. For the BIC in the passive structure without loss and gain, the Hermitian zero and pole coalesce at the real axis. The pole and zero correspond to the volume plasmon and Fabry-Perot modes, respectively. For the EP singularity, the non-Hermitian zero and pole coalesce at the real axis as well. The coalescence is associated with the simultaneous coherent perfect absorption (CPA) and lasing when transmission is simultaneously infinite and zero. This interpretation is confirmed by the sharp Fano profiles seen at $\theta_\pm$ in Fig. \ref{fig7}(c) with a very close dip (absorber) and peak (amplifier). Dip-peak pairs appear at the corresponding points of the singular borderline between the high-$\gamma_{EP}$ and low-$\gamma_{EP}$ regions in Fig. \ref{fig6} featuring the CPA-lasing effect. The point at $\omega_p$ and $\theta_{BIC}=\theta_b$ is a degenerate point (``BIC + CPA-lasing'') emerging due to the very peculiar conditions of the ENZ singularity and volume-plasmon excitation at the plasma frequency and, thus, having very special properties discussed in the rest of this paper. In particular, excitation of the BIC making the radiation to be mostly concentrated inside the structure also suppresses the CPA-lasing and results in the symmetric (Lorentzian) lineshape.

The coincidence of $\theta_{BIC}$ and $\theta_b$ has far-reaching consequences. In particular, just below the boundary, the $\mathcal{PT}$ symmetry can be broken by any loss and gain value, no matter how small. As a result, a sharp resonance appears in the place of BIC as shown in Figs. \ref{fig7}(b) and \ref{fig7}(c), so that the strict BIC transforms into the quasi-BIC. The resonance width reduces, when $\gamma$ decreases. The side resonances at $\theta_\pm$ which can be also associated with the border between regions in Fig. \ref{fig6}(a) clearly have the Fano profile with the regions of gain-assisted reflection above the unity especially pronounced in Fig. \ref{fig7}(c). The results remain essentially the same for the inverted structure ``gainy layer -- interlayer -- lossy layer'' (i.e., when light impinges the gainy layer) with the correction for the Fano profiles inversion.

We would like to emphasize that the quasi-BIC resonance at $\theta_{BIC}$ is characterized by the transmission which is close to the unity as shown in Fig. \ref{fig7}(d). In other words, this quasi-BIC is effectively free of both absorption and amplification. This property is kept when the non-Hermiticity magnitude increases: the resonance gets wider, but still close to the unity in the transmission peak. We explain this fact with the perfect loss and gain symmetry of the system in conditions of the BIC, what makes our system different from other examples of loss-induced quasi-BICs with resonant increase in absorption as in Refs. \cite{Sakotic2021, Ren2021}. On the contrary, the side resonances at $\theta_\pm$ have rapidly growing transmission and reflection when $\gamma$ increases. Note that we limit ourselves to relatively low realistic $\gamma$s, since for larger ones, $\gamma \sim 1$, the effects of instability (such as lasing) are able to violate the perfect transmission.

\begin{figure}[t!]
\centering \includegraphics[scale=1, clip=]{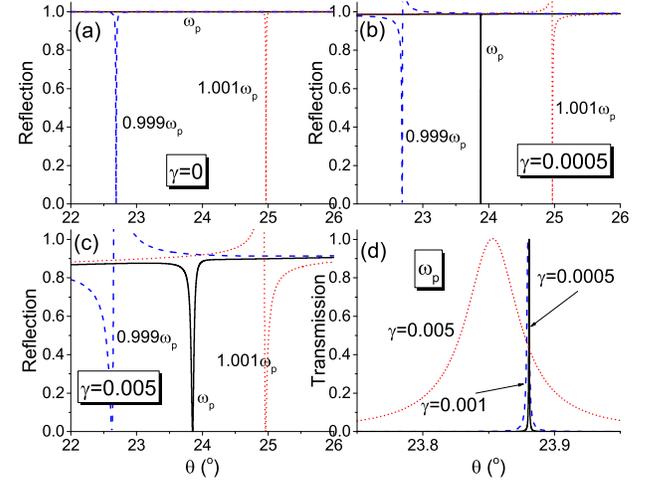}
\caption{\label{fig8} (a)-(c) Reflection and (d) transmission angular spectra of the $\mathcal{PT}$-symmetric trilayer for different loss and gain levels $\gamma$. The spectra at the frequencies $\omega_p$, $0.999 \omega_p$ and $1.001 \omega_p$ are shown. The other parameters are the same as in Fig. \ref{fig7}.}
\end{figure}

Figure \ref{fig8} shows the angular analog of Fig. \ref{fig7}. We see again the absence of the reflection dip at the plasma frequency for the lossless structure [Fig. \ref{fig8}(a)]. Introduction of the loss and gain transforms the BIC into the quasi-BIC at the singular point $\theta_{BIC}$ with the narrow resonance at $\omega_p$ and wider Fano resonances at other frequencies [Figs. \ref{fig8}(b) and \ref{fig8}(c)]. Finally, the transmission spectra in Fig. \ref{fig8}(d) demonstrate both widening and shift of the quasi-BIC resonance, when $\gamma$ increases, the peak transmission being almost equal to the unity. Note that the resonance shifts to the lower angles, that is into the low-$\gamma_{EP}$ region as in Fig. \ref{fig7}.

Symmetric shape of the resonances in Figs. \ref{fig7}(d) and \ref{fig8}(d) is a distinctive feature of some quasi-BICs as verified experimentally for individual nanoparticles \cite{Melik-Gaykazyan2021}. A small blue shift of the quasi-BIC resonance evident from Fig. \ref{fig7}(d) is caused by two reasons: (i) imperfect ENZ condition for a nonzero $\gamma$ and (ii) availability of low-$\gamma_{EP}$ values above $\omega_p$. We also emphasize the necessity of the balanced loss and gain for existing the quasi-BIC discussed, because there are no any resonances at $\theta_{BIC}$ for loss only as shown in Appendix \ref{AppE}. The sustained unity transmission resonance can be of interest for applications in tunable filtering and enhanced sensing, the tunability being caused by the influence of thickness and permittivity of the interlayer on the BIC position.

\begin{figure}[t!]
\centering \includegraphics[scale=1, clip=]{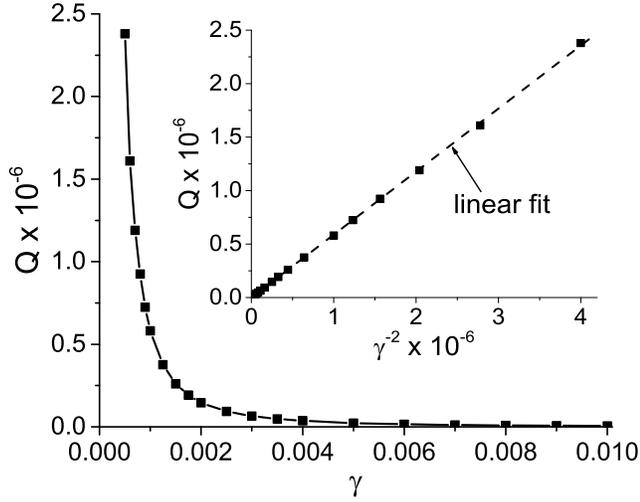}
\caption{\label{fig9} Dependence of the $Q$ factor of the quasi-BIC resonance on the loss and gain level $\gamma$. The $\mathcal{PT}$-symmetric trilayer is the same as in Fig. \ref{fig7}. The inset demonstrates the linear dependence of $Q$ on $\gamma^{-2}$.}
\end{figure}

The narrow quasi-BIC resonances discussed above possess extremely high $Q$ factors. Due to their symmetry, we can utilize a simple estimate, $Q=\omega_0/\Delta \omega$, where $\omega_0 \approx \omega_p$ is the resonance central frequency (we take into account its slight shift with increasing $\gamma$), $\Delta \omega$ is the resonance full-width at half maximum which can be directly estimated from the plots without any fitting. Figure \ref{fig9} demonstrates a sharp increase of the $Q$ factor when decreasing loss and gain level $\gamma$, so that $Q$ readily exceeds $10^6$ for $\gamma < 10^{-3}$. Moreover, the inset of Fig. \ref{fig9} demonstrates a linear dependence of the $Q$-factor on the inverse square of $\gamma$. Such a behavior is a well-known characteristic of the BIC violated by asymmetry \cite{Koshelev2018}. The asymmetry is often introduced to transform the exact unobservable BIC of the perfect structure to the observable quasi-BIC of the non-ideal system (e.g., the asymmetry may be due to a nonzero angle between elements of the structure as in Ref. \cite{Koshelev2018}). In our case, the non-Hermiticity magnitude $\gamma$ takes on the role of the structure asymmetry parameter, although the loss and gain are balanced and the shape of the quasi-BIC remains symmetric (so that the Fano asymmetry factor is infinite). This means that the asymmetry inducing the tranform of BIC into the quasi-BIC is caused exclusively by breaking $\mathcal{PT}$ symmetry.

High $Q$ factors correspond to the strong light localization inside the system as shown directly in our calculations of the intensity distribution inside the $\mathcal{PT}$-symmetric trilayer (Fig. \ref{fig10}) using the method described in Appendix \ref{AppF}. One can see that the quasi-BIC resonance is characterized by the symmetric intensity distribution due to loss compensation by gain [see Fig. \ref{fig10}(b)]. Note also that the peaks of the stationary interference pattern inside the interlayer have very high intensity in perfect accordance with the large value of the $Q$ factor. Detuning from $\omega_p$ results in asymmetric low-intensity distributions with attenuation in the loss layer uncompensated by the amplification in the gain layer [see Figs. \ref{fig10}(a) and \ref{fig10}(c)]. We would like to mention that the distributions almost do not change, if we swap the loss and gain layers.

\begin{figure}[t!]
\centering \includegraphics[scale=0.9, clip=]{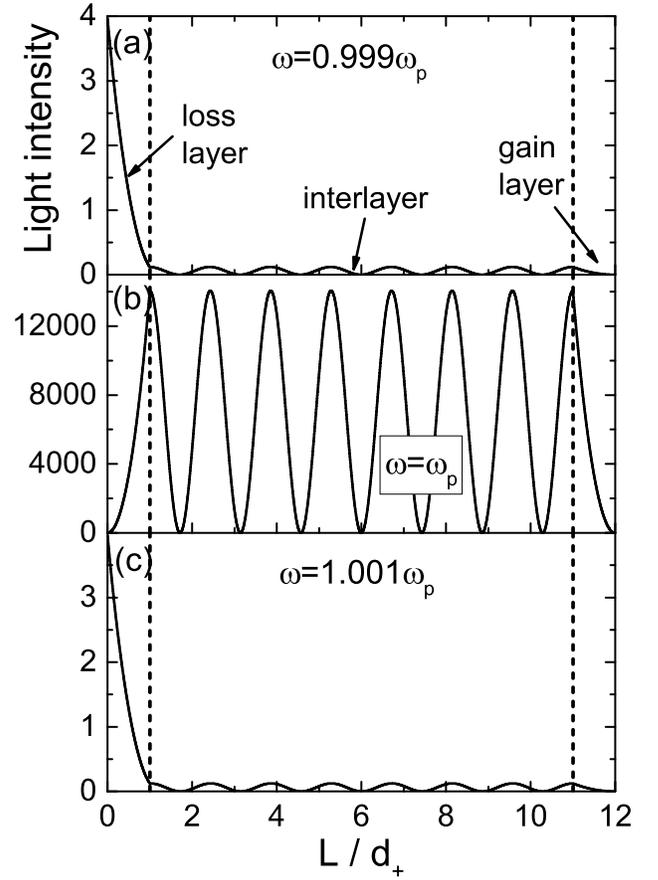}
\caption{\label{fig10} Normalized intensity distributions inside the $\mathcal{PT}$-symmetric trilayer for $\gamma=0.001$ and $\theta=\theta_{BIC}$ at the frequencies (a), (d) $0.999 \omega_p$, (b), (e) $\omega_p$, and (c), (f) $1.001 \omega_p$. The other parameters are the same as in Fig. \ref{fig7}.}
\end{figure}

\section{Conclusion} 

To sum up, we have studied violation of the $\mathcal{PT}$ symmetry in trilayer structures with balanced loss and gain. We prove that when the loss and gain layers are ENZ media, an extraordinary singular point of the $\mathcal{PT}$-symmetry-breaking phase diagram coincides with the BIC position enabling appearance of the high-$Q$ perfect-transmission resonance with the loss and gain value playing a role of the structure asymmetry parameter. We believe that this new way of quasi-BICs generation induced by $\mathcal{PT}$-symmetry breaking is of general interest and applicable in development of non-Hermitian photonics. To observe this effect in more complex structures supporting BICs, one has to tune conditions for the phase-diagram singularity of the $\mathcal{PT}$-symmetric system to match it with the BIC position of the same structure without loss and gain. From the different perspective, the poles and zeros of the lossless and $\mathcal{PT}$-symmetric structures should converge at the same point of the parameter space. In our case, this condition is fulfilled at the peculiar point of the ENZ (at the plasma frequency). For sophisticated photonic structures, this condition can be more intricate.

\acknowledgements{The work was supported by the Belarusian Republican Foundation for Fundamental Research (Project No. F20R-158) and the Russian Foundation for Basic Research (Project No. 20-52-00031).}

\appendix

\section{Basics of the transfer-matrix method \label{AppA}}

The transfer-matrix method is a convenient approach for calculation of stationary response of layered structures. We use it in the form presented in the Novotny and Hecht textbook \cite{Novotny}. Let us briefly describe the main points of this method, since they will be used in further derivations. Limiting ourselves to the TM-polarized plane waves, the relation between the amplitudes of the incident wave $e_0$, reflected wave $r$ and transmitted wave $t$ can be written as follows,
\begin{equation}
\left( \begin{array}{c} {e_0} \\ {r} \end{array} \right) = M \left( \begin{array}{c} {t} \\ {0} \end{array} \right).
\label{TMeq}
\end{equation}
The total transfer matrix $M=T_{01} \Phi_1 T_{12} \Phi_2 ... \Phi_n T_{n,n+1}$ is the product of the matrices $T_{i-1,i}$ taking into account light refraction at the interfaces between layers,
\begin{equation}
T_{i-1,i} = \frac{1}{2} \left( \begin{array}{cc} {1+\kappa_i \eta_i} & {1-\kappa_i \eta_i} \\ {1-\kappa_i \eta_i} & {1+\kappa_i \eta_i} \end{array} \right),
\label{Tmat}
\end{equation}
and the matrices $\Phi_i$ taking into account light propagation inside layers,
\begin{equation}
\Phi_i = \left( \begin{array}{cc} {e^{-i k_{i,z} d_i}} & {0} \\ {0} & {e^{i k_{i,z} d_i}} \end{array} \right).
\label{PHImat}
\end{equation}
Here $\kappa_i = k_{i-1,z}/k_{i,z} = \sqrt{(\varepsilon_{i-1} - \sin^2 \theta)/(\varepsilon_i - \sin^2 \theta)}$ is the ratio of longitudinal components of the wavevector in neighboring layers, $\eta_i=\varepsilon_i / \varepsilon_{i-1}$ is the ratio of the adjacent-layers permittivities, $\theta$ is the light incident angle, $d_i$ is the $i$th layer thickness. The $0$th and $(n+1)$th layers correspond to the semi-infinite ambient media, which we assume to be the air. Knowing the full transfer matrix of the structure $M$, one can easily compute the reflection and transmission coefficients normalized to the incident wave amplitude (i.e., $e_0=1$ is assumed) as $t=1/M_{11}$, $r_L=M_{21}/M_{11}$, and $r_R=-M_{12}/M_{11}$, where $M_{ij}$ is the corresponding component of the matrix $M$.

\section{Positions of different EPs \label{AppB}}

\begin{figure}[t!]
\centering \includegraphics[scale=1, clip=]{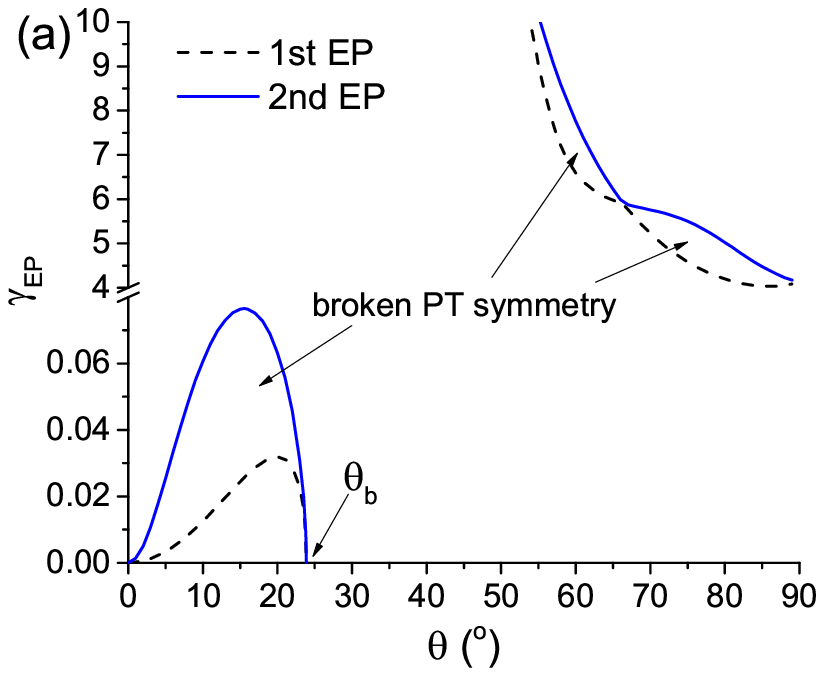}
\centering \includegraphics[scale=1, clip=]{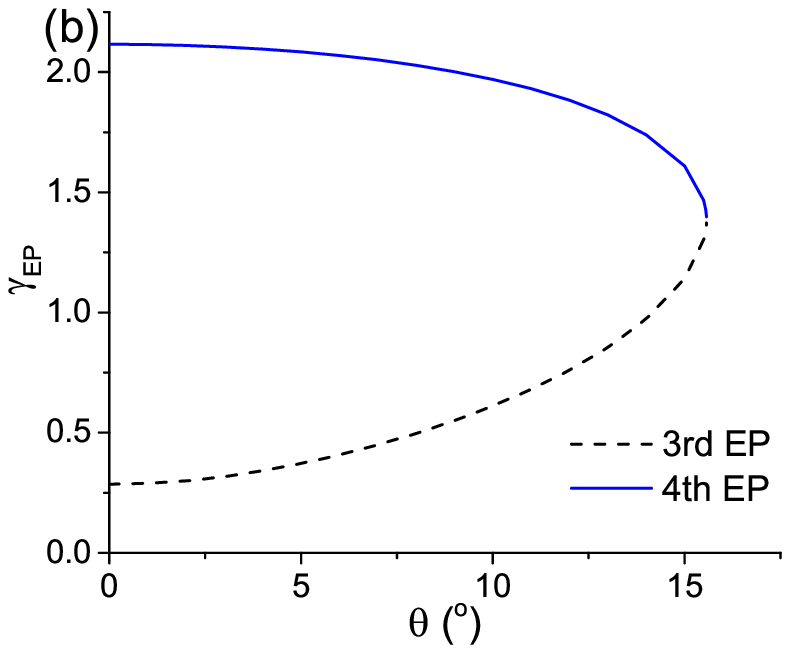}
\caption{\label{fig11} Angular dependencies of the positions of (a) the 1st and 2nd EPs and (b) 3rd and 4th EPs. The frequency is $\omega=\omega_p$; the other parameters are the same as in Fig. \ref{fig1}.}
\end{figure}

As we have seen in Fig. \ref{fig2}, the system may have several EPs at the same angle of incidence. In the main text, we have focused on the first EP as the most important one for us. In Fig. \ref{fig11}, we show change of positions of different EPs as a function of $\theta$. Regions of broken $\mathcal{PT}$ symmetry lie between the 1st and 2nd EPs as well as between the 3rd and 4th EPs. The latter region exists only at small incident angles. When approaching the singular point $\theta_b$, the region of broken $\mathcal{PT}$ symmetry between the 1st and 2nd EPs gets narrower.

\section{Derivation of the boundary angle value for the trilayer \label{AppC}}

The boundary angle $\theta_b$ (singular point) can be estimated from the condition of $|t| > 1$ for $\gamma \gg 1$. In terms of the transfer matrix, this means that $|M_{11}| < 1$ for $\gamma \gg 1$. For the three-layer structure discussed in the main text, we can give a relatively simple derivation of the transfer matrix at the plasma frequency. Indeed, for $\omega=\omega_p$ and $\gamma \gg 1$, the following relations are reduced to $\varepsilon_{\pm}=\pm i \gamma$; $\eta_1=i \gamma=-1/\eta_0$, $\eta_2=\varepsilon_{il}/i \gamma=-1/\eta_3$; $k_{1,z} \approx k_0 (1+i) \sqrt{\gamma/2}=-ik_{3,z}$, $k_{2,z}=k_0 \sqrt{\varepsilon_{il}-\sin^2 \theta}$; $\kappa_1 \approx \cos \theta/\sqrt{i \gamma}=i/\kappa_1$, $\kappa_2 \approx \sqrt{i \gamma/(\varepsilon_{il}-\sin^2 \theta)}=-i/\kappa_3$, where $k_0=\omega_p/c$. After some algebra, we obtain for the transfer matrix component of interest
\begin{equation}
|M_{11}| \approx 2 \gamma e^{\sqrt{2 \gamma}} \cos \theta \sqrt{1 - \frac{\sin^2 \theta}{\varepsilon_{il}}} \left| \sin \left( k_0 d_{il} \sqrt{\varepsilon_{il}-\sin^2 \theta} \right) \right|.
\label{M11}
\end{equation}
For arbitrarily large $\gamma$, this value remains limited only for arbitrarily small sine term. Thus, for $\gamma \rightarrow \infty$, the equation for the boundary angle $\theta_b$ reads
\begin{equation}
\sin \left( \frac{\omega_p}{c} d_{il} \sqrt{\varepsilon_{il}-\sin^2 \theta_b} \right) = 0,
\label{boundeq}
\end{equation}
which has the solution
\begin{equation}
\theta_b = \arcsin \sqrt{\varepsilon_{il}- \left( \frac{\pi c n}{\omega_p d_{il}} \right)^2},
\label{Sthetab}
\end{equation}
where $n=0,1,2,...$.

\section{EP position as a function of the interlayer thickness \label{AppD}}

\begin{figure}[t!]
\centering \includegraphics[scale=1, clip=]{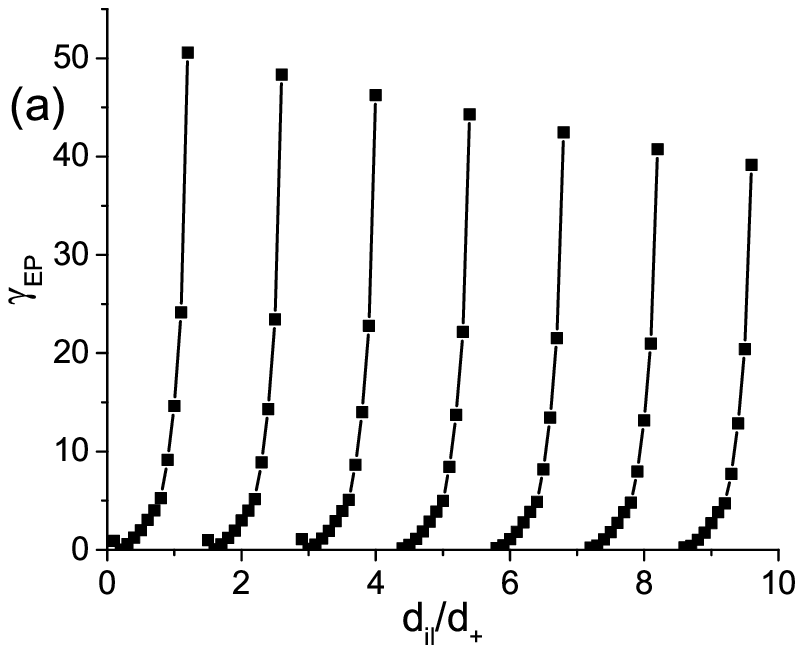}
\centering \includegraphics[scale=0.9, clip=]{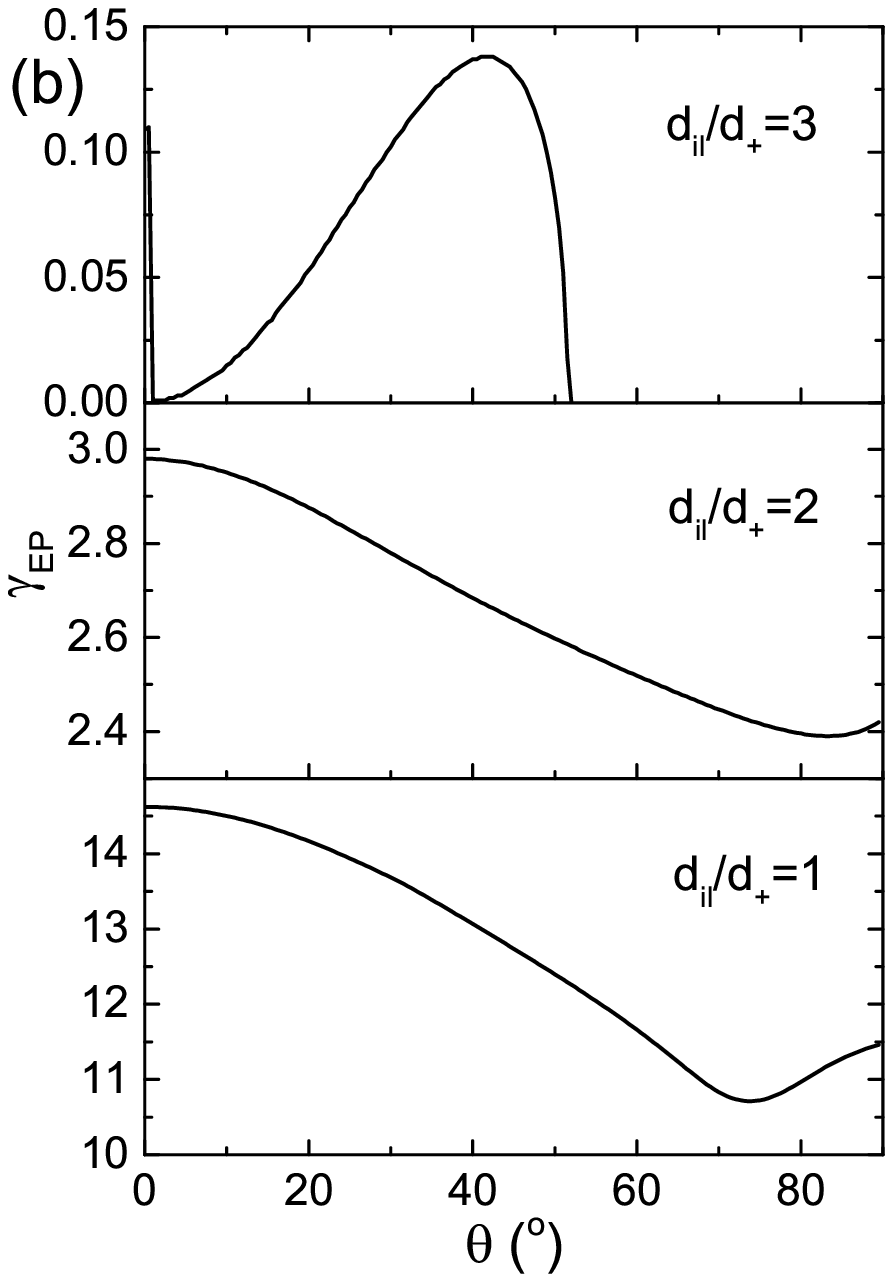}
\caption{\label{fig12} (a) The dependence of $\gamma_{EP}$ on the interlayer thickness at $\omega=\omega_p$ and $\theta=0$. (b) The angular dependence of the EP at different interlayer thicknesses at $\omega=\omega_p$. The other parameters are the same as in Fig. \ref{fig1}.}
\end{figure}

Here, we briefly discuss the influence of the interlayer thickness $d_{il}$ on the response of the structure. In the main text, we have considered mostly the case $d_{il}/d_\pm=10$ and seen the line of singular points in Fig. \ref{fig6} and the BIC at the plasma frequency. On the contrary, the bilayer ($d_{il}=0$) does not supports such features. In order to trace the transition between these two cases, we fix $\omega=\omega_p$ and $\theta=0$ and plot the EP position as a function of $d_{il}$ in Fig. \ref{fig12}(a). We observe a periodic dependence when regions of easy $\mathcal{PT}$-symmetry breaking take turns to the regions of tough $\mathcal{PT}$-symmetry breaking. Such a periodicity means that singular points are attainable not for every interlayer thickness, what is supported by Fig. \ref{fig12}(b): the singularity is seen at $d_{il}/d_\pm=3$, but not at $d_{il}/d_\pm=1$ or $2$. Thus, tuning the interlayer thickness is important for realizing the necessary regime of light interaction with the structure.

\section{The case of the purely absorbing ENZ media \label{AppE}}

\begin{figure}[t!]
\centering \includegraphics[scale=0.9, clip=]{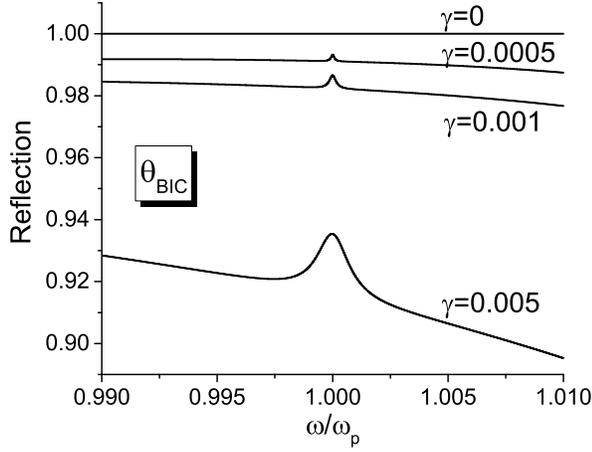}
\caption{\label{fig13} Reflection spectra of the purely lossy trilayer at the incident angle $\theta_{BIC}$ for different loss levels $\gamma$ of the first and third layers. The other parameters are the same as in Fig. \ref{fig7}.}
\end{figure}

Here we consider the case of purely lossy media used as a model of the ENZ layers. This means that the permittivities of the first and third layers of the trilayer are the same, being equal to $\varepsilon_\pm=1 + i \gamma - \omega^2_p/\omega^2$ with $\gamma>0$. Reflection spectra at the BIC angle $\theta_{BIC}$ for different $\gamma$ are shown in Fig. \ref{fig13}. We start from the strict BIC at $\gamma=0$ (there are no resonances). Introducing the loss we break the BIC. However, here the BIC breaks in a fundamentally different way in comparison to the case of balanced loss and gain considered in the main text. Loss causes a wide-band decrease of the reflection due to absorption. This results in the low-reflection background with a peak at the BIC position. In other words, the BIC resists losses and strives for its own preservation. On the contrary, in case of the balanced loss and gain, we see a high-reflection background with a sharp dip due to the easily broken BIC. We can make a conclusion that the loss itself is not the best way to transform a BIC into a quasi-BIC. The balance of the loss and gain is much more efficient.

\section{Calculation of intensity distributions inside the trilayer \label{AppF}}

In order to illustrate features of the perfect-transmission high-$Q$ quasi-BIC at $\omega=\omega_p$ and $\theta=\theta_{BIC}$, we calculate the distributions of the intensity inside the structure at the plasma frequency and neighboring frequencies with the approach described in Ref. \cite{Novitsky2008}. In particular, we divide the layers into many thin sublayers and utilize a partial transfer-matrix $M^{(i)}$ covering a part of the structure from its input interface to the $i$th sublayer as
\begin{equation}
\left( \begin{array}{c} {e_0} \\ {r_0} \end{array} \right) = M^{(i)} \left( \begin{array}{c} {t_i} \\ {r_i} \end{array} \right),
\label{partTMeq}
\end{equation}
where $r_0$ is the reflection coefficient of the entire structure, $t_i$ and $r_i$ are the amplitudes of the forward and backward waves in the $i$th sublayer. Then, we readily get
\begin{equation}
r_i=\frac{M^{(i)}_{11} r_0 - M^{(i)}_{21} e_0}{M^{(i)}_{11} M^{(i)}_{22} - M^{(i)}_{12} M^{(i)}_{21}}, \qquad
t_i=\frac{-M^{(i)}_{12} r_0 + M^{(i)}_{22} e_0}{M^{(i)}_{11} M^{(i)}_{22} - M^{(i)}_{12} M^{(i)}_{21}}.
\end{equation}
The normalized intensity inside a given sublayer is calculated as $I_i=|t_i+r_i|^2/e_0^2$.

\begin{figure}[b!]
\centering \includegraphics[scale=1., clip=]{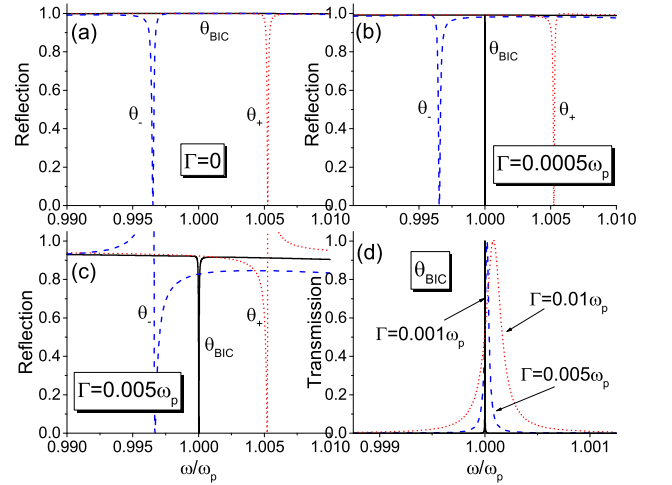}
\caption{\label{fig14} The same as in Fig. \ref{fig7}, but for the loss and gain layers described by the standard Drude formula.}
\end{figure}

\section{Modeling of the ENZ layers permittivity using the Drude model \label{AppG}}

Here we show that the results obtained in the main text with Eq. (\ref{epslg}) are in accordance with analogous calculations performed with the standard Drude formula for the ENZ media, $\varepsilon_\pm=1 - \omega^2_p/(\omega^2 \pm i \Gamma \omega)$. Figure \ref{fig14} corresponds to Fig. \ref{fig7}. Aside from minor changes in the positions of the Fano resonances, the features of the quasi-BICs induced by balanced loss and gain persist in the Drude-model case as well.

\end{document}